\documentclass[12pt]{article}
\usepackage{amsmath}
\def\cF{{\cal F}}
\def\cG{{\cal G}}
\def\cH{{\cal H}}
\def\cL{{\cal L}}
\def\cX{{\cal X}}

\def\a{\alpha} 
\def\b{\beta}
\def\d{\delta}          
\def\e{\eta}
\def\ga{\gamma}         
\def\la{\lambda}        
\def\ka{\kappa}
\def\m{\mu}
\def\n{\nu}

\def\s{\sigma}
\def\ee{\epsilon}

\def\idx{\int\! dt\, d^d\!x \,}

\def\pd{\partial}
\def\ds{\displaystyle}
\def\igual{\hspace{-7pt}=\hspace{-7pt}}

\def\menos{\hspace{-7pt}-\hspace{-7pt}}

\renewenvironment{itemize}
  {\begin{list}%
     {}%
     {\setlength{\topsep}{-1pt}%
      \setlength{\partopsep}{-6pt}%
      \setlength{\itemsep}{-3pt}%
      \setlength{\labelsep}{5pt}%
      \setlength{\itemindent}{0pt}%
     }%
  }%
  {\end{list}}%

\begin{document}
\begin{titlepage}

\rightline{FT-UCM 98-12/01}
\rightline{J. Phys. A Math. Gen {\bf 31} (1998) 8355}
\vskip 60pt
\begin{center}
     {\large {\bf Driven diffusion of vector fields}}\\
\vskip 40pt 
     {\rm S. Marculescu}\footnote{E-mail: 
          {\tt marculescu@physik.uni-siegen.d400.de}}
\vskip 4pt
     {\it Universit\"at-GH Siegen, D-57068 Siegen, Germany}
\vskip 25pt
     {\rm F. Ruiz Ruiz}\footnote{E-mail: 
          {\tt ferruiz@eucmos.sim.ucm.es}}\\
\vskip 4pt
     {\it Departamento de F\'{\i}sica Te\'orica I, 
          Facultad de Ciencias F\'{\i}sicas}\\
    {\it Universidad Complutense de Madrid, E-28040 Madrid, Spain}
\vskip 40pt
\begin{minipage}{390pt}
\noindent
A model for the diffusion of vector fields driven by external
forces is proposed. Using the renormalization group and the
$\epsilon$-expansion, the dynamical critical properties of the model
with gaussian noise for dimensions below the critical dimension are
investigated and new transport universality classes are obtained.
\end{minipage}
\end{center}
\end{titlepage}

\noindent 
Fourteen years ago it was suggested that driven diffusive systems may
provide suitable models for fast ionic conductors and solid
electrolytes \cite{Katz}. This triggered the interest in such systems,
which over the years have become a very good scenario to investigate
the properties of far from equilibrium steady states \cite{Zia}. A
simple microscopic realization of a driven diffusive system is an
Ising lattice gas in contact with a thermal bath in which an external
uniform driving force is introduced \cite{Katz}. The usual hopping
rate to nearest neighbours is modified through a bias in the rates for
hopping along the direction of the external field. In a finite lattice
with open boundary conditions, the gas evolves to a non-equilibrium
steady state with a non-zero particle current. Due to the energy
exchange of the system with the heat bath and the driving field, the
system also has a steady energy current.  So far driven diffusion of
scalar quantities, like the excess of transported particles, has been
considered. The purpose of this paper is to extend these
considerations to vector quantities.

We begin by very briefly recalling the formulation for the standard
case of a scalar quantity \cite{Janssen-1}. Space is taken to be
$d$-dimensional and parameterized by coordinates $(x_i)=
(x_\|,\mathbf{x}_\bot)$, with $\|$ the direction of the driving force
and $\bot$ the directions orthogonal to it. The deviation $s(x,t)$ of
the density from its uniform average satisfies a continuity equation
$\pd_t s + \pd_i J_i(s) = \e,$ with $\,J_i(s)\,$ the current generated
by ordinary diffusion and the drift in the direction of the driving
force, and $\eta(x,t)$ the noise accounting for the heat bath. The
contribution of ordinary diffusion to the current is proportional to
the gradient of $s$, while the drift in the $\|$-direction can be
estimated, for small density fluctuations $s$, by expanding in powers
of $s$ and retaining terms through order two. This gives the driven
diffusion equation \cite{Janssen-1}
\begin{equation}
    \pd_t s - \n \pd_\|^2 s 
       - \n_{\!\bot}\,\Delta_{\!\bot} s + \frac{\la}{2}\,\pd_\| s^2 
          = \eta~.
\label{DDS}
\end{equation}
In the first papers \cite{Janssen-1} the noise was assumed to be the
derivative of a random current, thus conserving the number of
particles. In subsequent papers \cite{Becker} gaussian noise
(describing systems with random particle sources and drains and
preserving particle number only in the mean) was also considered.  For
both types of noise it was shown that driven diffusive systems have,
for dimensions less than the critical dimension, infrared (IR) stable
fixed points at which various quantities scale. Furthermore the
corresponding critical exponents have been computed using the
$\ee$-expansion \cite{Janssen-1} \cite{Hwa} \cite{Becker}, the values
obtained showing that density fluctuations spread faster than
diffusively.

It is worth noting that eq.~(\ref{DDS}) also describes the toppling of
grains of sand in a sandpile \cite{Hwa}. As a point of fact, the
system described by eq.~(\ref{DDS}) with gaussian noise was first
considered as a model for the hydrodynamic regime of a running
sandpile. However one must mention that, after a shift, gaussian noise
has vanishing average and this requires particles to be randomly added
and removed, whereas in the running sandpile particles are only
added. In other words, gaussian noise does not reproduce the feeding
rate of the running sandpile \cite{Becker}. It is also interesting to
mention that for $d=1$ a field redefinition maps the driven diffusion
equation (\ref{DDS}) into the Kardar-Parisi-Zhang equation \cite{KPZ},
which in turn is identical \cite{Zia} to the Navier-Stokes
equation. This is no longer true for $d\neq 1$, for which the three
equations are different.

We now generalize driven diffusive systems to the case where the
external forces acting on the system produce drift in a hyperplane of
dimension $n$ parameterized by coordinates $x_\a ~(\a=1,\ldots,n)$. We
do this by increasing the number of field variables from 1 to $n$, so
that we have a vector $s_\a(x,t)$. The component $s_\a(x,t)$ describes
the excess of grains transported along the direction $x_\a$. For every
component $s_\a(x,t)$ we write a continuity equation $\pd_t s_\a +
\pd_i J_{i\a} = \e_\a$. Proceeding now as for the the standard case,
we obtain
\begin{equation}
   \pd_t s_\a - \s \pd _\b \pd_\b s_\a - 2\s\xi\,\pd_\a\pd_\b s_\b 
              - \n_{\!\bot} \Delta_{\!\bot} s_\a 
              + \frac{\la}{2} \> \pd_\a s^2
              + \ka \,\pd_\b \big( n s_\a s_\b 
              - \d_{\a\b} s^2\big) = \e_\a ~ ,
\label{DDS-n}
\end{equation}
where we have used the notation $s^2\equiv s_\a s_\a$. This equation
has two anisotropy coefficients, $\s$ and $\s\xi$, and two coupling
constants, $\la$ and $\ka$. For later convenience, we introduce the
coefficient
\begin{equation}
   \n\equiv \s(1+2\xi)\>. 
\label{nu}
\end{equation}
Eq.~(\ref{DDS-n}) makes sense for $1\leq n \leq d$. If $n=1$, we
recover the standard driven diffusion equation (\ref{DDS}); if $n=d$,
anisotropy is maximal and $\n_{\!\bot}=0$. We want to investigate the
dynamical critical properties of driven diffusive systems based on
this equation. In this paper, we concentrate on gaussian noise
\begin{equation}
\begin{array}{c}
  \langle \e_\a(x,t)\rangle = 0 \\[9pt]
  \langle \e_\a(x,t)\,\e_\b(x',t')\rangle = 2 \>\d_{\a\b} \>
      \d^{(d)}(x-x')\>\d(t-t')
\end{array}
\label{gaussian}
\end{equation}
and use the renormalization group and the $\ee$-expansion to
characterize to first order in $\ee$ the universality classes of the
model for dimensions below the critical dimension. A similar program
can be carried for conserving noise.

We have already mentioned that eq.~(\ref{DDS-n}) may be used to
describe drift in a hyperplane. Scenarios in which this may become
relevant are oil diffusion and diffusion in disorder media. More
generally, eq.~(\ref{DDS-n}) may be considered in connection with
directed percolation \cite{Stauffer}. We recall that in directed
percolation, when the probability $p$ of jumping to the nearest
neighbour along a preferred direction is set equal to a critical value
$p_*$, a phase transition occurs.  At this transition, the system
shows scaling behaviour together with non-standard features, like {\it
e.g.}  fractal dimensionality for the percolating cluster. One could
think of the parameter $\xi$ as playing the role of the probability
$p$, with the difference that $\xi$ is not {\it a priori} set equal to
any critical value but enters the model as one other parameter. The
continuity equation (\ref{DDS-n}) can also be thought of as describing
the toppling of grains of sand in an $n$-dimensional sandpile, with
$s_\a(x,t)$ the deviation in the direction $x_\a$ of the sandpile
profile from its uniform average. Since in general $n\neq 1$, one may
think of using eq.~(\ref{DDS-n}) to describe the hydrodynamic regime
of sandpiles with avalanches advancing anisotropically. Yet we must
keep in mind that gaussian noise does not describe the feeding rate of
a sandpile. Other Langevin equations based on gaussian noise
describing surface sandpiles have been proposed in ref. \cite
{Mehta}. They differ from (\ref{DDS-n}) in that they double the number
of field variables but keep the number of transport directions equal
to one. It is important to realize that eq.~(\ref{DDS-n}) is not the
spherical generalization of the ordinary driven diffusion equation
(\ref{DDS}) in which, while keeping the number of anisotropy
directions equal to one, $s(x,t)$ becomes an $n$-vector field
$(n\!=\!2j\!+\!1)$ carrying an $n$-dimensional irreducible unitary
representation of $O(3)$. For isotropic spherical generalizations of
the Navier-Stokes equation and the Kardar-Parisi-Zhan equation, see
refs. \cite{You} and \cite{Doherty}.

To study the system (\ref{DDS-n})+(\ref{gaussian}), we use
ref. \cite{Janssen-2} and recast it as a stochastic quantum field
theory model with classical action 
\begin{equation}
\begin{array}{l}
   {\ds S_n = \idx \,\pi_\a \bigg[\!- \pi_\a+ \pd_t s_\a   
              - \s \pd _\b \pd_\b s_\a - 2\s\xi\,\pd_\a\pd_\b s_\b 
              - \n_{\!\bot} \Delta_{\!\bot} s_\a }\\[3pt]
   \phantom{\ds S_n =\idx \,\pi_\a \bigg[ }
      {\ds +\, \frac{\la}{2} \> \pd_\a s^2
        + \ka \,\pd_\b \big( n s_\a s_\b 
        - \d_{\a\b} s^2\big)\,\bigg] ~,} 
\end{array}
\label{action}
\end{equation}
where $\pi_\a(x,t)$ is the response field \cite{MSR} to the assumed
gaussian noise. The action $S_n$ has the following symmetries:
\begin{itemize}
\item[(a)] Translational invariance in all coordinates, 
\item[(b)] $O(n) \otimes O(d\!-\!n)\,$ invariance with respect to
spatial coordinates, 
\item[(c)] Joint reflection of longitudinal coordinates, field
variables and response fields, 
\item[(d)] Scaling of longitudinal coordinates
$(t,x_\a,\mathbf{x}_\bot)$ $\to$ $(t,\varrho x_\a,\mathbf{x}_\bot)$,
with the anisotropy coefficients, coupling constants and fields
transforming as $\,(\s,\xi,\n_{\!\bot}) \to (\varrho^2
\s,\xi,\n_{\!\bot}),$ $\,(\la,\ka) \to \varrho^{(n+2)/2}\,(\la,\ka)\,$
and $\,(s_\a,\pi_\a) \to \varrho^{-n/2}\,(s_\a,\pi_\a)$. Since the
parameter $\xi$ is scale invariant, we introduce the invariant
coupling constants
\begin{displaymath}
   u \equiv \frac{1}{4\pi} \> \frac{\la}{\s^{(n+2)/4}} \qquad\qquad
   \qquad v \equiv \frac{1}{4\pi} \> \frac{\ka}{\s^{(n+2)/4}} ~. 
\end{displaymath}
\end{itemize}
In addition, if $n=1$, the action is invariant under Galilei
transformations \cite{Becker} \cite{Hwa}
\begin{displaymath}
\begin{array}{l}
  s(x,t)\to s(x_{\bot},x_\|-\la at,t)+ a \\[3pt]
  \pi(x,t) \to \pi(x_{\bot},x_\|-\la at,t) ~.
\end{array}
\end{displaymath}
We note that this symmetry does not generalize to $n>1$. The reason is
that Galilei's invariance requires a constant velocity vector and this
is only possible for $n=1$, since in this case there is only one
preferred direction. In other words, it is precisely the passage from
$n=1$ to $n\neq 1$ what destroys the Galilei symmetry.

The critical dimension of the model is $d_c=4$. For $d=4$ the model is
ultraviolet renormalizable by power counting, the only primitively
divergent one-particle irreducible Green functions being the
self-energy $\langle \pi_\a(p) s_\b(-p) \rangle$ and the vertex
$\langle \pi_\a(p) s_\b(q) s_\ga(-p-q)\rangle$, which have linear and
logarithmic degrees of divergence respectively. Using dimensional
regularization \cite{tHooft}, we obtain the following one-loop
renormalization group equations:
\begin{eqnarray} 
   {\ds \b_u \equiv \frac{du}{d\ln\m} } &\igual& 
      {\ds - u \bigg( \frac{\ee}{2} + \cH + n\cL 
                    + \frac{n+2}{4}\>\cF \bigg) } \label{RG1}\\[6pt] 
   {\ds \b_v \equiv \frac{dv}{d\ln\m} } &\igual& 
      {\ds - v \bigg( \frac{\ee}{2} + \cH + \frac{n+2}{4}\>\cF \bigg) } 
                                                 \label{RG2}\\[6pt] 
   {\ds \b_\xi \equiv \frac{d\xi}{d\ln\m} } &\igual& 
      {\ds \xi\,(\cG - \cF) } \label{RG3}\\[6pt]
   {\ds \zeta_\s \equiv \frac{d\ln\s}{d\ln\m} } &\igual& {\ds \cF }
                                                 \label{RG4}\\[6pt] 
   {\ds \zeta_\n \equiv \frac{d\ln\n}{d\ln\m}} &\igual& 
      {\ds \frac{\cF +2\xi\cG}{1+2\xi} ~,} \label{RG5}
\end{eqnarray}
where $\m$ is the dimensional regularization mass scale, $\ee=4-d$ is
the deviation from the critical dimension and $\cF,\,\cG,\,\cH$ and
$\cL$ are functions of $u,\,v$ and $\xi$ given by
\begin{displaymath}
    n\,(n+2)\,\cX(u,v,\xi) = u^2 x_{uu}(\xi) + uv\, x_{uv}(\xi) 
                           + v^2 x_{vv}(\xi)
    \qquad \cX \equiv \cF, \cG, \cH, \cL~.
\end{displaymath}
For completeness we write the expressions of $x_{uu},\,x_{uv}$ and
$x_{vv}$:
\begin{eqnarray*}
   f_{uu} &\igual& 0 \\
   f_{uv} &\igual& {\ds \bigg[\, \frac{n^2}{\xi} 
           - n\, (n+1)\,(n-2)\,\bigg]\, a_1 
           - n^2\,\bigg(\frac{1}{\xi} + 2\bigg)\,a_2
           - \frac{n }{2} ~ (n+4)\,a_3 }\\[6pt]
   f_{vv} &\igual& {\ds \frac{n^2}{2}~(-2n^2+n+4) 
           - 2n\,\bigg( \frac{n}{\xi} +n + 2\bigg)\,a_1 
           + n^2\,\bigg(\frac{n+2 }{\xi}+4\bigg)\,a_2  
           - n^2\,\bigg( \frac{n}{\xi}-n+2 \bigg)\, a_3}\\
            &\menos& {\ds n\,(n^2-n-4)\, a_4 }\\[12pt]
   g_{uu} &\igual& {\ds -\,\frac{n+2}{2\xi}~\bigg[\, (n-1)\> a_1 
             + \bigg(1-\frac{n}{4}\bigg) \, a_4 \bigg] }\\[6pt]
   g_{uv} &\igual& {\ds \frac{n}{2 \xi} ~\bigg[\!\!
               - \frac{1}{4}\> (n-1)\,(n+2)\,(3n+4) 
           + \bigg( \frac{n-2}{\xi}
                         - n^2+5n+2-\frac{8}{n} \bigg)\,a_1 }\\[3pt] 
          &\menos&{\ds (n-2)\,\bigg(\frac{1}{\xi}+2\bigg) \,a_2  
           + (n-1)\,(n+2)\>a_3 
           - \bigg( \frac{3}{2}\>n^2+4n-\frac{8}{n} 
                 \bigg)\, a_4 \bigg] }
\end{eqnarray*}
\begin{eqnarray*}
  \hphantom{ f_{vv} } &\hphantom{\igual}& 
    \hphantom{  {\ds \frac{n^2}{2}~(-2n^2+n+2) 
           - 2n\,\bigg( \frac{n}{\xi} +n + 2\bigg)\,a_1 
           + n^2\,\bigg(\frac{n+2 }{\xi}+4\bigg)\,a_2  
           - n^2\,\bigg( \frac{n}{\xi}-n+2 \bigg)\, a_3}}\\[-24pt]
   g_{vv} &\igual& {\ds \frac{n-2}{2n\xi}~f_{vv} }\\[12pt]
   h_{uu} &\igual& {\ds \frac{2}{\xi}~[ \, (n+1)\> a_1
             - n a_2- a_3 ] }\\[6pt]
   h_{uv} &\igual& {\ds \frac{2n}{\xi}~ \bigg[\, n^2-2 
           - (n+1)\,\bigg(n-2+\frac{4}{n}\bigg)\,a_1 
           - 2\,(n-2)\,a_2
             + \bigg(n-2+\frac{4}{n}\bigg)\, a_3 + 2a_4 \bigg]
             }\\[6pt]             
   h_{vv} &\igual& {\ds -\,\frac{n}{\xi} ~\bigg[\, 4\,(n^2-2)  
           + \bigg( n^3-4n^2+2n-\frac{8}{n}\bigg) \,a_1 
           + 2\,(n-2)^2\, a_2 }\\[3pt]
          &\menos& {\ds \bigg( n^3+2n^2-6n+8-\frac{8}{n}\bigg)\,a_3
             + 8 a_4 \bigg] }\\[12pt]
   \ell_{uu} &\igual& {\ds \frac{1}{\xi}~[- (n+3)\,a_1 
           + 2 a_2 + (n+1)\>a_3\,] }\\[6pt]
   \ell_{uv} &\igual& {\ds \frac{2n}{\xi}~\bigg[\, 1 
           + \frac{6}{n}\>a_1 
           + \frac{2\,(n-2)}{n}\>a_2 
           + \bigg(n-2-\frac{2}{n}\bigg)\,a_3 
             - (n+1)\>a_4 \bigg] }\\[6pt]
   \ell_{vv} &\igual& {\ds -\,\frac{n}{\xi}~\bigg[\, 4 
           + \bigg(n-4+\frac{12}{n}\bigg)\, a_1
           - \frac{2\,(n-2)^2}{n}\> a_2 
           + \bigg(5n -4-\frac{4}{n}\bigg)\,a_3 
             - 4\,(n+1)\,a_4 \bigg]~, }
\end{eqnarray*}
with $a_1,\,a_2,\,a_3$ and $a_4$ taking the form
\begin{displaymath}
\begin{array}{ll}
  {\ds a_1 = a_2\> F\bigg(\frac{n}{2},1;2;\frac{\xi}{1+\xi} \bigg)}  
     &  {\ds a_2 = \frac{1}{(1+\xi)^{n/2}} }\\[12pt] 
  {\ds a_3 = a_4 \> F\bigg(\frac{n}{2},1;2;-\frac{\xi}{1+\xi}\bigg) 
     \qquad\qquad} & {\ds a_4 = \frac{1}{(1+2\xi)^{n/2}} } 
\end{array} 
\end{displaymath} 
and $F(a,b;c;z)$ being the hypergeometric function. For $n=1$ there is
no $\ka$-term in $S_n$, so that $\ka=0$ and hence $v=0$. From this,
and using $\,n=1$, it follows that $\,\cF=\cH + n\cL=0$. Thus
$\b_u=\ee/2$ and, as a result, $\b_u=0$ at the critical dimension,
implying no ultraviolet renormalization of $\la$. This lack of
renormalization holds to all orders in perturbation theory and follows
\cite{Hwa} from the non-anomalous character of the Galilei symmetry of
the $n=1$ model mentioned above. For an $\,n=1\,$ driven diffusive
system which is classically Galilei invariant but has a quantum
mechanical anomaly, see ref. \cite{Janssen-anomaly}.

To characterize the critical behaviour of the model for $d\leq d_c$,
we solve the renormalization group equations in the neighbourhood of
all IR attractive fixed points
$(u_*,v_*,\xi_*)$. Eqs.~(\ref{RG1})-(\ref{RG3}) form a system of
differential equations describing the dependence of $u,\,v$ and $\xi$
on $\m$. Its solution near an IR attractive fixed point gives, upon
substitution in eqs.~(\ref{RG4})-(\ref{RG5}), the $\m$-dependence of
the anisotropy coefficients $\s$ and $\n$ near the fixed
point. Indeed, in the neighbourhood of a fixed point the solutions to
eqs.~(\ref{RG4})-(\ref{RG5}) read $\,\s\propto \m^{\zeta_{\s *}}\,$
and $\,\n\propto \m^{\zeta_{\n *}}\,$, with
\begin{displaymath} 
  \zeta_{\s *} \equiv {\cF\vert}_* \qquad\qquad\qquad 
  \zeta_{\n *} \equiv {\frac{\cF+2\xi\cG}{1+2\xi}\bigg\vert}_* ~.
\end{displaymath}
To have a signal for criticality, at least one of the two exponents
$\zeta_{\s *}$ and $\zeta_{\n *}$ must be negative, since $\s$ or $\n$
will then diverge as $\m\to 0$. To find the IR attractive fixed point
with diverging length scales of the model for $d\leq d_c$, we
therefore proceed in two steps:

\noindent
{\it Step 1.} We first find all fixed points by solving the equations
$\b_u=\b_v=\b_\xi=0$ and retain those which are IR attractive. We
remind that for a fixed point to be IR attractive, the matrix of
derivatives
\begin{equation}
  \left( \begin{array}{ccc}
       \beta_{uu} ~ & ~ \beta_{uv} ~ & ~ \beta_{u\xi} \\
       \beta_{vu} ~ & ~ \beta_{vv} ~ & ~ \beta_{v\xi} \\
       \beta_{\xi u} ~ & ~ \beta_{\xi v} ~ & ~ \beta_{\xi\xi}
         \end{array} \right)
  \qquad \beta_{xy} = \frac{\partial \b_x}{\partial y}\>\bigg\vert_* 
\label{linear}
\end{equation}
that results from linearizing eqs.~(\ref{RG1})-(\ref{RG3}) about the
fixed point must have eigenvalues with strictly positive real
parts. The case $\xi_*\to\infty$ must be treated separately, since
then there is no renormalization of $\xi$, hence no equation
(\ref{RG3}). In this case, fixed points are solutions of
$\,\b_u=\b_v=0\,$ and the matrix of derivatives of the beta functions
is the upper-left $2\!\times\! 2$ submatrix of (\ref{linear}).

\noindent
{\it Step 2.} Next we calculate $\zeta_{\s *}$ and $\zeta_{\n *}$ for
all the IR attractive fixed points found in step 1 and keep those
values which are negative. For $\xi_*$ finite, the condition
$\,\b_\xi=0\,$ and eq.~(\ref{RG3}) imply that the two critical
exponents take the same value:\footnote{This also follows from the
definitions of $\n,~ \zeta_\s$ and $\zeta_n$ given in eqs. (\ref{nu}),
(\ref{RG4}) and (\ref{RG5}).} $\,\zeta_{\s *}=\zeta_{\n *}$. For
$\,\xi_*\to\infty$, however, the condition $\,\b_\xi=0\,$ is no longer
required and $\zeta_{\s *}$ and $\zeta_{\n *}$ may take different
values.

Although $n$ has been assumed to take integer values in the range 1 to
4, the results can be analytically continued in $n$.  In fact,
analytic continuation gives well-defined beta functions for exotic
values of $n$, like for example $n\to 0$ or $n\to -2$. In what follows
we take $n$ as a real variable.

In step 1 of this program, one must keep in mind that the values $u_*$
and $v_*$ of the couplings $u$ and $v$ at a fixed point are not
physically meaningful. Hence redefining $\bar{u}=\chi u$ and
$\bar{v}=\psi v$, with $\chi$ and $\psi$ functions of $\xi$, does not
change the physics. Such a redefinition may be used when one or both
of $u_*$ and $v_*$ go to infinity to move the fixed point to
$\bar{u}_*$ and $\bar{v}_*$ finite. The renormalization group
equations for $\bar{u}$ and $\bar{v}$ are straightforward to obtain
from eqs.(\ref{RG1})-(\ref{RG3}) and involve the functions $\chi$ and
$\psi$. The functions $\chi$ and $\psi$ must be found such that the IR
attractive fixed points of the new renormalization group equations
have $\bar{u}_*$ and $\bar{v}_*$ finite. This introduces two new
unknowns, $\chi$ and $\psi$, which, together with the complexity of
the functions $\cF,\,\cG,\,\cH$ and $\cL$, makes it not feasible to
analytically carry the program above while keeping $\xi$ and $n$
arbitrary. We have thus restricted ourselves to some values of $n$ and
$\xi$ that we find of interest. In particular, we have treated the
cases 
\begin{itemize}
\item[(i)] $n=-2,0,1,2,3,4$ with arbitrary $(u_*,v_*,\xi_*)$, and 
\item[(ii)] $\xi_*=0,\,(n-2)/2n,\,\infty$ with arbitrary $(n,u_*,v_*)$,
\end{itemize}
allowing in both instances for redefinitions of $u$ and $v$ as
described above. The values $\xi_*=0$ and $\xi_*=\infty$ considered in
case (ii) correspond respectively to $\pi_\a\pd^2s_\a$ and
$\pi_\a\pd_\a\pd_\b s_\b$ being the dominant kinetic anisotropic term
in the classical action $S_n$. The idea of looking for critical points
at $\,\xi_*=(n-2)/2n\,$ stems from the simple relation between
$f_{vv}$ and $g_{vv}$ for arbitrary $n$, namely $2n\xi
g_{vv}=(n-2)f_{vv}$. Because of this relation, the equation $\b_\xi=0$
simplifies considerably if $\,\xi_*=(n-2)/2n$, thus making possible to
find IR attractive fixed points with finite $\xi_*$. Another
motivation to consider $\,\xi_*=(n-2)/2n\,$ is to investigate whether
the passage from $n\neq 1$ to $n=1$ is smooth, since the case $n=1$
can be recovered from $\xi_*=-1/2$ and $\s\to\infty$. 

The results that we have obtained for (i) and (ii) are as follows (we
provide the values of the physical quantities $n,~ \zeta_{\n *}$ and
$\zeta_{\s *}$, which characterize the universality classes). 
\begin{table}[ht]
\begin{center}
\begin{tabular}{|c|c|c|c|}
\hline
   $n$ & $\xi_*$ & $\zeta_{\n *}$ & $\zeta_{\s *}$  \\
\hline 
   3 & \begin{tabular}{c} 1/6 \\ $\infty$ \\ $\infty$ 
       \end{tabular} 
     & \begin{tabular}{c} $-0.1257\ee$ \\ positive \\ $-0.1068\ee$
       \end{tabular} 
     & \begin{tabular}{c} $-0.1257\ee$ \\ $-0.2848\ee$ \\ 
        $-0.4726\ee$ \end{tabular} \\
\hline 
   4 & \begin{tabular}{c} 1/4 \\ $\infty$ \\ $\infty$ 
       \end{tabular} 
     & \begin{tabular}{c} $-158\ee/629$ \\ positive \\ $-0.3731\ee$
       \end{tabular} 
     & \begin{tabular}{c} $-158\ee/629$ \\ $-0.1324\ee$ \\ 
        $-0.3526\ee$\end{tabular} \\
\hline
\end{tabular} 
\\[9pt]
{\sl Table 1: Universality classes for $n=3,4$.}
\end{center}
\vspace{-15pt}
\end{table}
For case (i) there are no IR attractive fixed points with diverging
length scales if $n=-2,0,2$. If $n=1$, the results of ref. \cite{Hwa}
are recovered. And if $n=3,4,$ we obtain the universality classes in
Table 1. For case (ii) we obtain the universality classes:

$\bullet ~ \xi_*=(n-2)/2n$, with $n>2$ and critical exponents
\begin{equation}
   \zeta_{\s *}=\zeta_{\n *} 
      = -\, {\frac{ 2f_{vv}\,\ee }{
         (n+2)\,f_{vv} +4 h_{vv}}~\bigg\vert}_{\xi=(n-2)/2n} ~.
\label{crit-1}
\end{equation}
Note that the critical exponent is a complicated but well-defined
function of $n$. 

$\bullet ~ \xi_*=\infty,$ with $\,n=n_0\equiv(1+\sqrt{33})/4$ and 
critical exponents
\begin{equation}
  \zeta_{\n *} =0 \qquad \zeta_{\s *}= - \ee~.
\label{crit-2}
\end{equation}

$\bullet ~\xi_*=\infty,$ with $n>2$ and critical exponents
\begin{eqnarray}
  \zeta_{\s *}^\pm &\igual & \frac{1}{2}\> n b_1 \ee  
     \frac{ 2\,(n+1)\,X_\pm + n\,b_3 }{
          X_\pm\,(b_1 b_5 - b_2 b_4) + b_1 b_6 -b_3 b_4}
     \label{crit-3a} \\[9pt]
  \zeta_{\n *}^\pm &\igual& \frac{1}{16}\,(n-1)\,\ee   
     \frac{ X_\pm\,[\,n\,(3n+4)\,(n-2)\,b_1 - 8b_2\,] - 8b_3 
          }{X_\pm\,(b_1 b_5 - b_2 b_4) + b_1 b_6 -b_3 b_4} ~,
   \label{crit-3b}
\end{eqnarray}
where 
\begin{displaymath}
\begin{array}{ll}
  {\ds b_1 = -\,\frac{1}{n-2}~(3n^2 + 7n -2)} 
         & {\ds b_4 = -\,\frac{1}{n-2}~(n^2-3n-6) } \\[9pt]
  {\ds b_2 = -\,\frac{1}{8}~ n\,(3n+1)\,(n^2+2n-12)} \qquad
         & {\ds b_5 = \frac{1}{2}~ n\,(3n^2-3n-10) } \\[9pt] 
  {\ds b_3 = \frac{1}{2}~(2n^2-n -4) }
         & {\ds b_6 = -\,\frac{1}{8}~n^2\,(n+2)\,(2n^2-n-4)}
\end{array}
\end{displaymath}
and 
\begin{displaymath}
   X_\pm = \frac{1}{2 b_1} \,(-b_2 \pm \sqrt{b_2^2-4b_1 b_3}) ~.
\end{displaymath}
The exponents $\zeta_{\n *}^-$ and $\zeta_{\s *}^\pm$ are negative for
all $\,2<n\leq 4$, whereas $\,\zeta_{\n *}^+\,$ is negative only for
$\,2<n<2.0060$.  It is worth noting that the universality classes for
$\,n=-2,0,2,3,4\,$ are particular cases of
eqs.~(\ref{crit-1})-(\ref{crit-3b}).

The anomalous dimensions $\zeta_{\n *}$ and $\zeta_{\s *}$ found above
are related to power laws in physical correlation functions and
therefore should be directly measurable. In accordance with the
continuity equation (\ref{DDS-n}) and owing to the vector character of
$s_\a$, physical correlators are expectation values of products of
tensor currents
\begin{displaymath}
  J_{\a\b} = -\s\,\pd_\a s_\b - 2\s\xi\,\pd_\b s_\a 
           + \frac{\la}{2}~ \d_{a\b} s^2 
           + \ka\,(n s_\a s_\b - \d_{\a\b} s^2)~. 
\end{displaymath}
Since $J_{\a\b}$ contains linear and quadratic terms in $s_\a$, the
computation of physical correlators involves renormalization of
composite operators. For $n=1$, due to the non-renormalization of the
coupling constant, renormalization of composite operators can be fully
eliminated by considering correlators quadratic in the current
\cite{Becker}. For $\,n\neq 1$, since the coupling constants do
renormalize, correlators cubic in $J_{\a\b}$ must also be
considered. The analysis of physical correlators and their relation to
the critical exponents $\zeta_{\n *}$ and $\zeta_{\s *}$ is in
progress and will be presented elsewhere.

We conclude with a few comments:

{\it Comment 1}. The anisotropy dimension $n$ at which the critical
exponents $\,\zeta_{\n *}\,$ and $\,\zeta_{\s *}$ occur is in general
non-integer. It is particularly noticeable that for $\,\bar{u}^2_*=0$,
$\,\bar{v}^2_*\,$ finite and $\,\xi_*=\infty$, the only IR attractive
fixed point with diverging length scales is that given in
eq.~(\ref{crit-2}) and occurs at $n=(1+\sqrt{33})/4$. With the caution
due to the one-loop nature of our results, this suggests a fractal
behaviour of the model.  The similarity in some regards of our model
with directed percolation may provide some insight in understanding
these points.

{\it Comment 2}. The fixed points associated with the critical
exponents in eqs.~(\ref{crit-1})-(\ref{crit-2}) have either
$\,\bar{u}_*=0\,$ or $\,\bar{v}_*=0$, while the fixed points
associated with the critical exponents in
eq.~(\ref{crit-3a})-(\ref{crit-3b}) have $\,\bar{u}_*,\bar{v}_*\neq
0$. We have not been able to find IR attractive fixed points with all
three parameters $\bar{u}_*,\,\bar{v}_*$ and $\xi_*$ finite and
non-zero. Our results show that if such a point exists to first order
in $\ee$, it has fractal dimension. 

{\it Comment 3}. Eqs.~(\ref{crit-3a})-(\ref{crit-3b}) give for $n>2$
two sets of critical exponents at $\xi_*=\infty$ which are associated
with two different attractive IR fixed points. This is so since
$\,\bar{u}^2_*\,$ and $\,\bar{v}^2_*\,$ are two-valued functions of
$\xi$ for $\,\xi\to\infty\,$ and $\,\bar{u}_*,\bar{v}_*\neq
0$. Depending on the configuration of all fixed points, the system
will flow in parameter space to one or another IR fixed point and will
reach one or another set of critical exponents.

{\it Comment 4}. If we take $\la=0$ in $S_n$, we find universality
classes at $\,\xi_*=\infty\,$ for $\,n>n_0$, with critical exponents
$\,\zeta_{\n *}=0\,$ and $\,\zeta_{\s *} = -2\ee/(n+2)$. On the other
hand, if we set $\,\ka=0$, we obtain universality classes at
$\,\xi_*=\infty\,$ for $\,n>2$, with critical exponents $\,\zeta_{\n
*}=0\,$ and $\,\zeta_{\s *} = -\ee/6$. Note however that, despite the
simplicity and beauty of these results, for $\la=0$ or $\ka=0$
perturbation theory may generate at higher loops divergent relevant
operators which are not in the classical action $S_n$. For this
reason, we have considered the more general case of two coupling
constants.

\vspace{25pt}

{\bf Acknowledgment.} The authors are grateful to Bo Zheng for many
clarifying discussions. FRR is grateful to the Institute for
Theoretical Physics in Heidelberg for its hospitality and to the
Alexander von Humboldt Foundation for support through a Research
Fellowship.

\end{document}